\documentclass{prop2015}% no class options needed by now
\usepackage[english]{babel}
\title{The Massless Spectrum of Heterotic Compactifications}
\author[Xenia de la Ossa]{Xenia de la Ossa\inst{1}}
\author[Edward Hardy]{Edward Hardy\inst{2}}
\author[ Eirik Eik Svanes]{Eirik Eik Svanes\inst{3,4,5,}\footnote{Corresponding author\quad E-mail:~\textsf{esvanes@lpthe.jussieu.fr}}}

\address[1]{Mathematical Institute, University of Oxford, Andrew Wiles Building, Radcliffe Observatory Quarter, Woodstock Road, Oxford OX2 6GG, UK}
\address[2]{Abdus Salam International Centre for Theoretical Physics, Strada Costiera 11, 34151, Trieste, Italy}
\address[3]{Sorbonne Universit\'es, UPMC Univ Paris 06, UMR 7589, LPTHE, F-75005, Paris, France}
\address[4]{CNRS, UMR 7589, LPTHE, F-75005, Paris, France}
\address[5]{Sorbonne Universit\'es, Institut Lagrange de Paris, 98 bis Bd Arago, 75014 Paris, France\\}

\shortauthors{X. de la Ossa et al.}
\begin{abstract}
We discuss the four-dimensional massless spectrum of supersymmetric Minkowski compactifications of ten-dimensional heterotic supergravity, including the anomaly cancelation condition. This can be calculated from restrictions arising from F-term conditions in a four dimensional effective theory. The results agree with computations of the infinitesimal moduli space recently performed from a ten-dimensional perspective. The paper is based on a talk given by Eirik Eik Svanes in Leuven for the workshop on 'The String Theory Universe'.
 \end{abstract}
\shortabstract
\begin{document}
\maketitle
%%% Use this if the article text won't start with a \section:
% \noindent
%%% Being based on LaTeX's article class, and2012 supports the respective 
%%% sectioning level from \section to \subparagraph.

\section{The Superpotential}
For a compactification of ten-dimensional heterotic supergravity to four dimensions to preserve $N=1$ supersymmetry, the internal space must have an $SU(3)$-structure $\left(X,\Omega,\omega\right)$, where $\omega$ is the hermitian two-form (K\"ahler form), and $\Omega\in\Omega^{(3,0)}(X)$ is a complex top-form that encodes the complex structure. The superpotential of the effective theory theory is then given by \cite{Gukov:1999gr, Becker:2003yv, Cardoso:2003af}
\begin{equation}
W= \int_X(H+i {\textrm{d}}\omega)  \wedge\Omega\: .
\end{equation}
Here 
\begin{equation}
H={\rm d}B+\frac{\alpha'}{4}\left(\omega_{CS}(A)-\omega_{CS}(\nabla)\right)\:,
\end{equation}
where $H_0 = {\textrm{d}}B$ for a two-form potential $B$, and $\omega_{CS}$ are the Chern-Simons forms. The F-term conditions for unbroken supersymmetry are
\begin{equation}
\delta W=W=0\,.
\end{equation}
These are equivalent to requiring \cite{Strominger1986253}
\begin{align}
{\rm d}\Omega & =0 \\
F^{(0,2)}&=R^{(0,2)}=0\\ 
H&=i(\partial-\bar\partial)\omega\:,
\end{align}
where $F$ and $R$ are the curvatures of $A$ and $\nabla$. That is, $X$ is complex, and $\left(V,A\right)$ and $\left(TX,\nabla\right)$ are holomorphic bundles.

Supersymmetry also requires that $X$ is conformally balanced. The remaining supersymmetry conditions requires that the bundles satisfy the Yang-Mills condition, and can be derived from D-terms in the effective four-dimensional theory \cite{Anderson:2009nt}. For the present paper, we will assume that these conditions are satisfied, in particular that the bundles are stable \cite{Yau87}. See however \cite{delaOssa:2014cia, delaOssa:2015maa} for more details on this.

\section{Infinitesimal Moduli: The Massless Spectrum}
The massless spectrum of the four-dimensional theory can be obtained from computing the infinitesimal moduli space of the the ten-dimensional solution \cite{Anderson:2014xha, delaOssa:2014cia}, or equivalently from the effective theory as we shall show. Further details may be found in \cite{delaOssa:2015maa}. At the supersymmetric locus, the mass-matrix of the four dimensional effective theory is given by
\begin{equation}
V_{I\bar J}=e^\mathcal{K}\partial_I\partial_KW\partial_{\bar J}\partial_{\bar L}\bar W\mathcal{K}^{K\bar L}\:,
\end{equation}
where $\mathcal{K}$ is the K\"{a}hler potential.  To obtain the massless moduli we must therefore find the deformations $\delta_2$ for which
\begin{equation}
\delta_2\delta_1W=0\;\;\;\;\forall\;\;\delta_1\:.
\end{equation}
where $\delta_1$ is a generic deformation. Physically this is demanding that $\delta_1W$ which is a generic F-term is not generated by deforming in the direction $\delta_2$. It follows that
\begin{equation}
{\textrm{d}}\delta_2\Omega =0\;\;\Rightarrow\;\;\delta_2\Omega \in H^{(2,1)}(X)\;\;\Leftrightarrow\;\;\Delta_2 \in H^{(0,1)}(TX)\:,
\end{equation}
where $\Delta_2$ is the complex structure deformation corresponding to $\delta_2\Omega$. We also get that
\begin{align}
{\mathcal{F}}(\Delta_2)&=\Delta_2^a\wedge F_{a\bar b}\,{\rm d}z^{\bar b}={\bar\partial}\alpha_2\\
{\mathcal{R}}(\Delta_2)&=\Delta_2^a\wedge R_{a\bar b}\,{\rm d}z^{\bar b}={\bar\partial}\kappa_2\:,
\end{align}
where $\alpha_2\in\Omega^{(0,1)}({\textrm{End}}(V))$, $\kappa_2\in\Omega^{(0,1)}({\textrm{End}}(V))$, whose closed part are bundle moduli. This implies that $\Delta_2$ is in the kernel of Atiyah maps \cite{MR0086359,Anderson:2011ty} 
\begin{align}
\mathcal{F}\:&:\;\;\; H^{(0,1)}(TX)\rightarrow H^{(0,2)}({\textrm{End}}(V))\\
\mathcal{R}\:&:\;\;\; H^{(0,1)}(TX)\rightarrow H^{(0,2)}({\textrm{End}}(TX))\:.
\end{align}
We can think of $x_2=\Delta_2+\alpha_2+\kappa_2$ as the moduli of a holomorphic structure
\begin{equation}
{\bar\partial}_1 ={\bar\partial}+{\mathcal{F}}+{\mathcal{R}}
\end{equation}
defining an extension sequence
\begin{equation}
0\rightarrow{\mathbf g}\rightarrow\mathcal{Q}_1\rightarrow TX\rightarrow0\:,
\end{equation}
where ${\mathbf g}={\rm End(TX)}\oplus{\rm End(V)}$. Note that ${\bar\partial}_1^2 =0$ iff the Bianchi identities: $\bar\partial{\mathcal{F}}={\bar\partial}{\mathcal{R}}=0$ are satisfied. The infinitesimal moduli space is \cite{MR0086359,Anderson:2011ty}
\begin{equation}
T{\mathcal{M}}_1=H^1(\mathcal{Q}_1)=H^1({\mathbf g})\oplus\ker({\mathcal{F}}+{\mathcal{R}})\:.
\end{equation}
It should be noted that $H^1({\rm End}(TX))\subseteq H^1({\mathbf g})$ are unphysical, but may be viewed as infinitesimal ${\mathcal{O}}(\alpha')$ field redefinitions \cite{delaOssa:2014msa}.

Assuming that $H^{(0,1)}(X)=0$ or the $\partial\bar\partial$-lemma, the final constraint from the condition $\delta_2\delta_1 W=0$ is that
%\footnote{To get \eqref{eq:FinalddW} we assume that $H^{(0,1)}(X)=0$.}
\begin{align}
\label{eq:FinalddW}
{\mathcal{H}}(x_2)&=  2\Delta^a_2\wedge i{\partial}_{[a}\omega_{b]\bar c}{\rm d}z^{b\bar c}  -\frac{{\alpha'}}{2}({\textrm{tr}}\:\alpha_2\wedge F-{\textrm{tr}}\:\kappa_2\wedge R)\notag\\
&={\bar\partial} y_2^{(1,1)}\:,
\end{align}
for some $y_2^{(1,1)}$. The closed part of $y_2^{(1,1)}$ gives the hermitian moduli. It follows that $x_2\in T{\mathcal{M}}_1$ is in the kernel of the map
\begin{equation}
{\mathcal{H}}\::\;\;\;H^1({\mathcal Q}_1)\rightarrow H^{(0,2)}(T^*X)\:.
\end{equation}
Again, we can think of $z_2=x_2+y_2$ as moduli of a holomorphic structure
\begin{equation}
{\bar\partial}_2={\bar\partial}_1+{\mathcal{H}}\:,
\end{equation}
defining the extension sequence
\begin{equation}
0\rightarrow T^*X\rightarrow\mathcal{Q}_2\rightarrow\mathcal{Q}_1\rightarrow0\:.
\end{equation}
Note again that ${\bar\partial}_2^2=0$ iff the heterotic Bianchi identity
\begin{equation}
{\rm d} H=\frac{\alpha'}{4}({\rm tr}\,F^2-{\rm tr}\,R^2)
\end{equation}
is satisfied. Overall the massless spectrum is given by \cite{Anderson:2014xha,delaOssa:2014cia}
\begin{equation}
T{\mathcal{M}}_2=H^1(\mathcal{Q}_2)=H^{(0,1)}(T^*X)\oplus\ker({\mathcal{H}})\:,
\end{equation}
where $H^{(0,1)}(T^*X)$ are the hermitian moduli.

%\bibliographystyle{prop2015}
%\bibliography{bibliography}

%%% Use the following two code lines if you wish to generate your bibliography with BibTeX;
%%% please replace first the string "demo" below with the name(s) of
%%% the BibTeX data base(s) you want to use.
%%% The resulting bibliography-output (the contents of the .bbl file)
%%% must be pasted into this file before submission.
%%% 
%%% \bibliographystyle{prop2015}
%%% \bibliography{demo}
%%% 
%%% If you are doing it by hand make sure to put the correct make up into 
%%% it.  Please see below for the usual keys; and please keep in mind that all the custom 
%%% macros (\jr, \othercit} and remarkably also \textsc  here have no effect 
%%% regarding typesetting.  They are crucial for the hypertext markup of the 
%%% data, though.
%%% 
%%% The macros are:
%%% \textsc for authors' names;  also for editors' names, if there are no authors;
%%% \jr for (abbreviated) journal names;
%%% \othercit to be used as a prefix to \bibitem in non-journal entries and like 
%%% a common makro for partial entries in multi-entry \bibref-constructs.
%%% 
%%% Replace the following example bibliography with your references
%%% before submission:

\end{document}